\documentclass{aa}
\usepackage{graphics}

\begin{document}

\thesaurus{09.04.1,10.07.01,11.13.1,11.09.4,11.19.5}

\title{Background galaxies as reddening probes throughout the Magellanic Clouds}
 
\author{C.M. Dutra\inst{1,3},  E. Bica\inst{1,3}, J.J. Clari\'a\inst{2,3}, A.E. Piatti\inst{2,3} \and A.V. Ahumada\inst{2,3} }

\offprints{C.M. Dutra (dutra@if.ufrgs.br)}

\institute{ Instituto de Fisica-UFRGS, CP 15051, CEP 91501-970 POA - RS, Brazil \and Observatorio Astron\'omico de C\'ordoba, Laprida 854, 5000, C\'ordoba, Argentina \and Visiting Astronomer, Complejo Astron\'omico  El Leoncito operated under agreement between the Consejo Nacional de Investigaciones Cient\'{\i}ficas y T\'ecnicas de la Rep\'ublica Argentina and the National Universities of La Plata, C\'ordoba and San Juan.}

\titlerunning{Reddening towards the Magellanic Clouds}
\authorrunning{Dutra et al.}

\date{Received; accepted}

\maketitle

\begin{abstract} 
We study the spectral properties in the range 3600{\rm  \AA} \ - 6800{\rm \AA} \ of the nuclear region of galaxies behind the Magellanic Clouds. The radial velocities clarified the nature of the objects as background galaxies or extended  objects belonging to the Clouds. For most galaxies behind the main bodies of the LMC and SMC, radial velocities were measured for the first time. In the present sample typical LMC background galaxies are nearby (4000 $< V({\rm km/s}) <$ 6000), while SMC's are considerably more distant (10000 $< V({\rm km/s}) <$ 20000). We determine the reddening in each line of sight by matching a reddening-free galaxy template with comparable stellar population. For the LMC main body we derive a combined Milky Way and internal reddening value E(B-V)$_{MW+i}$ = 0.12$\pm$0.10, while for the SMC E(B-V)$_{MW+i}$ = 0.05$\pm$0.05. By subtracting Milky Way reddening values for galaxies projected onto the surroundings of each Cloud, we estimate average internal reddening values $\Delta$E(B-V)$_i$ = 0.06 and 0.04, respectively for the main bodies of the LMC and SMC. The Clouds are optically thin, at least in the directions of the studied background galaxies which are often difficult to be identified as such on ESO/SERC sky survey images. Nevertheless, more reddened zones may occur where it is difficult to identify galaxies.

\end{abstract} 
 
 
\section{INTRODUCTION} 
 
The galactic reddening distribution was initially modeled as a function 
of galactic latitude in terms of a Cosecant Law (Sandage 1973 and references therein). A dependence on galactic longitude was derived by de 
Vaucouleurs et al. (1976).  The polar cap zero points have been a matter of debate with estimates ranging from reddening free to E(B-V)=0.05. These models do not take into account
local reddening variations, since the dust distribution can be very patchy, including high galactic latitudes where many discrete clouds occur (e.g. Reach et al. 1998). The empirical formulation
 by Burstein \& Heiles (1978, 1982) improved the reddening distribution description by relating H\,I column density
 and galaxy counts to reddening. Nevertheless the H\,I and dust contents do not scale in the same way everywhere, in particular in cold dense clouds where hydrogen becomes mostly molecular. 
 
Recently, Schlegel et al. (1998, hereafter SFD98) provided a new 
estimator of galactic reddening by means of a full-sky 100 {\rm $\mu$m} 
IRAS/ISSA map which was converted to dust column density by using 
a dust colour temperature map (17{\rm $^{\circ}$K} to 21{\rm $^{\circ}$K}) derived from 100 and 240 {\rm $\mu$m} 
COBE/DIRBE maps. The dust emission map is 
calibrated in terms of E(B-V) reddening values using determinations from early type galaxies by means of the (B-V) {\it vs} Mg2 
relation. SFD98's dust emission reddening E(B-V)$_{FIR}$ appears to be sensitive to the dust content of the cold dense clouds and their accumulation in different lines of sight. Dutra \& Bica (2000) 
compared E(B-V)$_{FIR}$ values with the reddening E(B-V) measured from the stellar content of globular and old open clusters in the Galaxy. It was concluded that differences between these reddening values most probably  arise from dust distribution in the cluster foreground and 
background.

 The stellar content of background galaxies offers an opportunity to analyse the total Milky Way dust column in a given line of sight, as well as those of the Magellanic Clouds. Recently the NED database included a facility tool to determine  SFD98's E(B-V)$_{FIR}$ in any direction. However for the main bodies  of the LMC and SMC NED adopted uniform foreground reddening values of E(B-V)$_{FIR}$ = 0.075 and E(B-V)$_{FIR}$ = 0.037, respectively. These values are from the average dust emission in their surroundings (SFD98). We emphasize that in the present study we use SFD98's original facility tool (dust-getval.f) and the whole sky dust emission reddening maps, which include the internal dust emission structure of the LMC and SMC E(B-V)$_{FIR}$ maps. As pointed out by SFD98 they have not analysed the E(B-V)$_{FIR}$ values in the Clouds.

The Magellanic Clouds cover a significant portion of the sky and 
their background galaxies have not yet been studied in detail due in part to reddening and crowding effects. Oestreicher et al. (1995) mapped the galactic 
reddening in the direction of the LMC by 
means of UBV photometry of foreground galactic stars. They 
obtained a mean reddening of E(B-V)$_{MW}$ = 0.06$\pm$0.02. For the SMC the mean foreground reddening is E(B-V)$_{MW} $$\approx$ 0.03 considering 
colour-magnitude diagrams of clusters in the outer parts of the SMC such as 
K3, L1 and NGC121 (Westerlund 1990). Oestreicher \& Schmidt-Kaler 
(1996) studied the internal reddening distribution of the LMC with dust clouds in the range  0.04  $<$ E(B-V)$_i$ $<$ 0.40. Reddening estimates using background galaxies by means of count 
methods have been applied for the SMC (Wesselink 1961, Hodge 1974, 
MacGillivray 1975) and LMC (Gurwell \& Hodge 1990). These studies 
provide very high  reddening values in some regions which seem to arise from count losses owing to star crowded fields and extended objects belonging to the Clouds. 
 
 In the present study, we observe and analyse  integrated spectra of nuclear regions of galaxies behind the main bodies of the Clouds to probe reddening in those lines of sight. We take into account different stellar populations by using template spectra. For comparison purposes we also observed galaxy spectra in the surroundings of the Clouds and towards the Galactic South Polar Cap. In Sect. 2 we present the samples in the directions of the Clouds and South Polar Cap. In Sect. 3 we describe the observations and reductions. In Sect. 4 we establish the nature of the objects. Most of them turn out to be galaxies, but some are extended objects belonging to the Clouds. In Sect. 5 we compare the present galaxy spectra with those of stellar population templates from Bica (1988, hereafter B88), and provide  some new templates as well. In Sect. 6 we derive the reddening value for each galaxy behind the Magellanic Clouds and discuss their  distribution and overall tranparency of the Clouds. Concluding remarks are given in Sect. 7.

\section{The samples}

In the main bodies of the Magellanic Clouds it is often difficult to establish the morphological type of a galaxy using ESO/SERC Schmidt plates and Digitized Sky Survey atlases. In some cases is not possible to distinguish a galaxy from an extended object belonging to the Clouds, such as a compact H\,II region or star cluster.
 
\subsection{Objects towards the Magellanic Clouds} 
 
We selected galaxies behind the Magellanic Clouds and also included objects with uncertain classification and/or excluded from the revised and extended catalogues of extended objects in the SMC (Bica \& Schmitt 1995, Bica \& Dutra 2000) and LMC (Bica et al. 1999) to establish their nature.  These LMC or SMC main body objects are: (i) SMC-DEM92(AM0054-744ne), LMC-DEM225 and LMC-DEM329 from the SMC and LMC catalogues of emission nebulae by Davies et al. (1976); (ii) HS75-8, HS75-10, HS75-13, HS75-18, HS75-20, HS75-22, HS75-23 and HS75-25 from the catalogue of galaxies behind the SMC by Hodge \& Snow (1975); (iii) HS17, HS45, HS257, HS356, HS394, HS449 and HS451 from the LMC star cluster catalogue by Hodge \& Sexton (1966); (iv) SL887 from the LMC star cluster catalogue by Shapley \& Lindsay (1963); (v) OHSC3 from the LMC star cluster catalogue  by Olszewski et al. (1988); and (vi) HW60 from the SMC star cluster catalogue by Hodge \& Wright (1974).

Figures 1 and 2 show the angular distribution of the observed objects towards the SMC and LMC respectively. We distinguish the literature galaxies that we observed (LEDA and/or NED extragalactic databases) from the objects whose nature will be established in the present study. The latter objects are excellent reddening probes for the central regions of the Clouds. A preliminary discussion of part of the Magellanic Clouds sample was given in Dutra et al. (1998).

The 43 selected objects are presented in Table 1, by columns: (1) designation, (2) and (3) J2000 equatorial coordinates, (4) and (5) galactic coordinates, (6) total magnitude B$_T$ (LEDA/NED), (7) exposure time, (8) radial velocity measurement, (9) LEDA and/or NED radial velocity, and (10) LEDA/NED  morphological type or estimated by ourselves on ESO/SERC plates when possible. Additional objects included in Table 1 are discussed subsequently.

\begin{figure*}
\resizebox{\hsize}{!}{\includegraphics{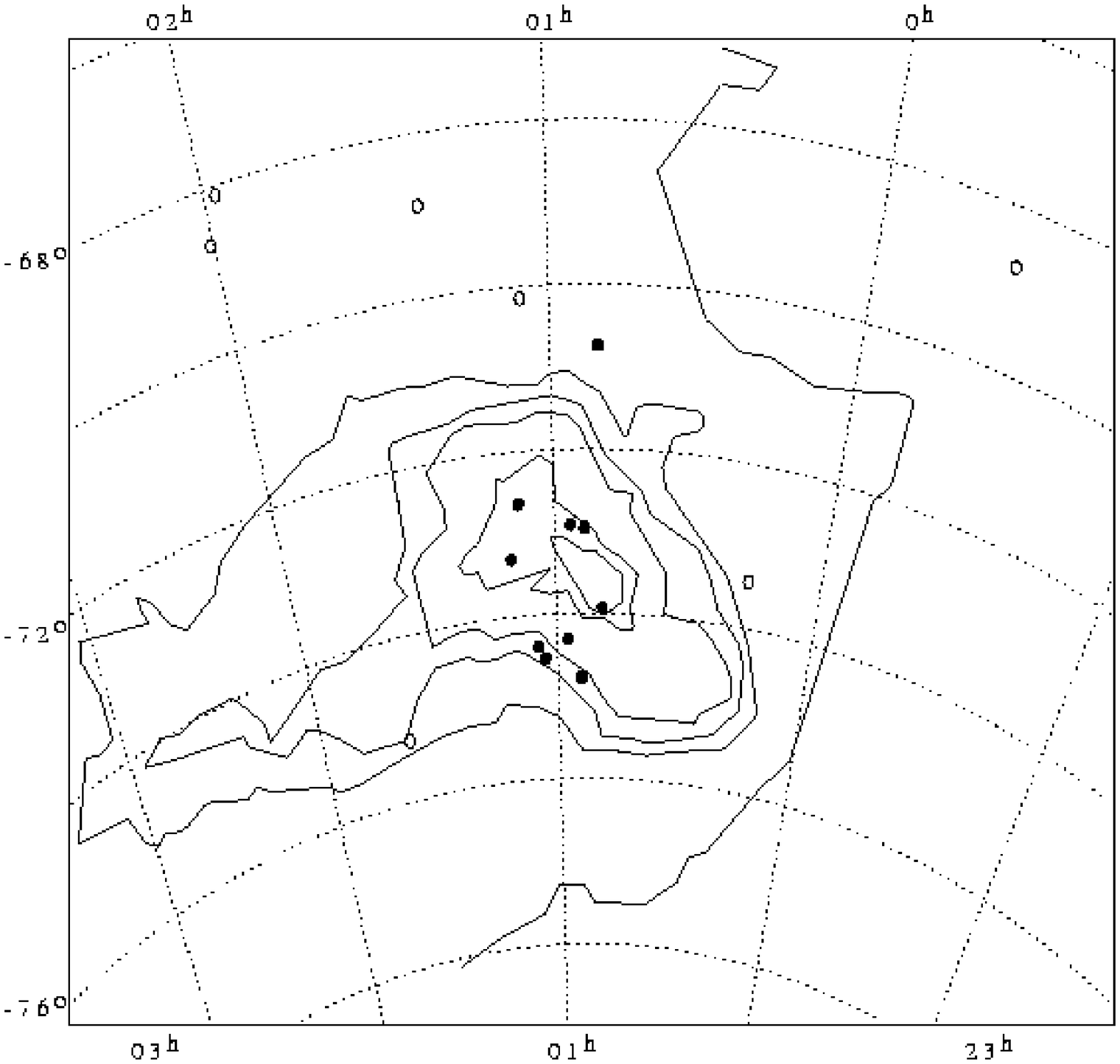}}
\caption{Angular distribution of the observed SMC background known galaxies (open circles) and candidate galaxies (filled circles). The solid lines represent H\,I contours of 5, 50, 100, 150, 400 and 600 in units {\rm $10^{19} atoms$ $ cm^{-2}$} from Mathewson \& Ford (1984).}
\label{sample}
\end{figure*}

\begin{figure*}
\resizebox{\hsize}{!}{\includegraphics{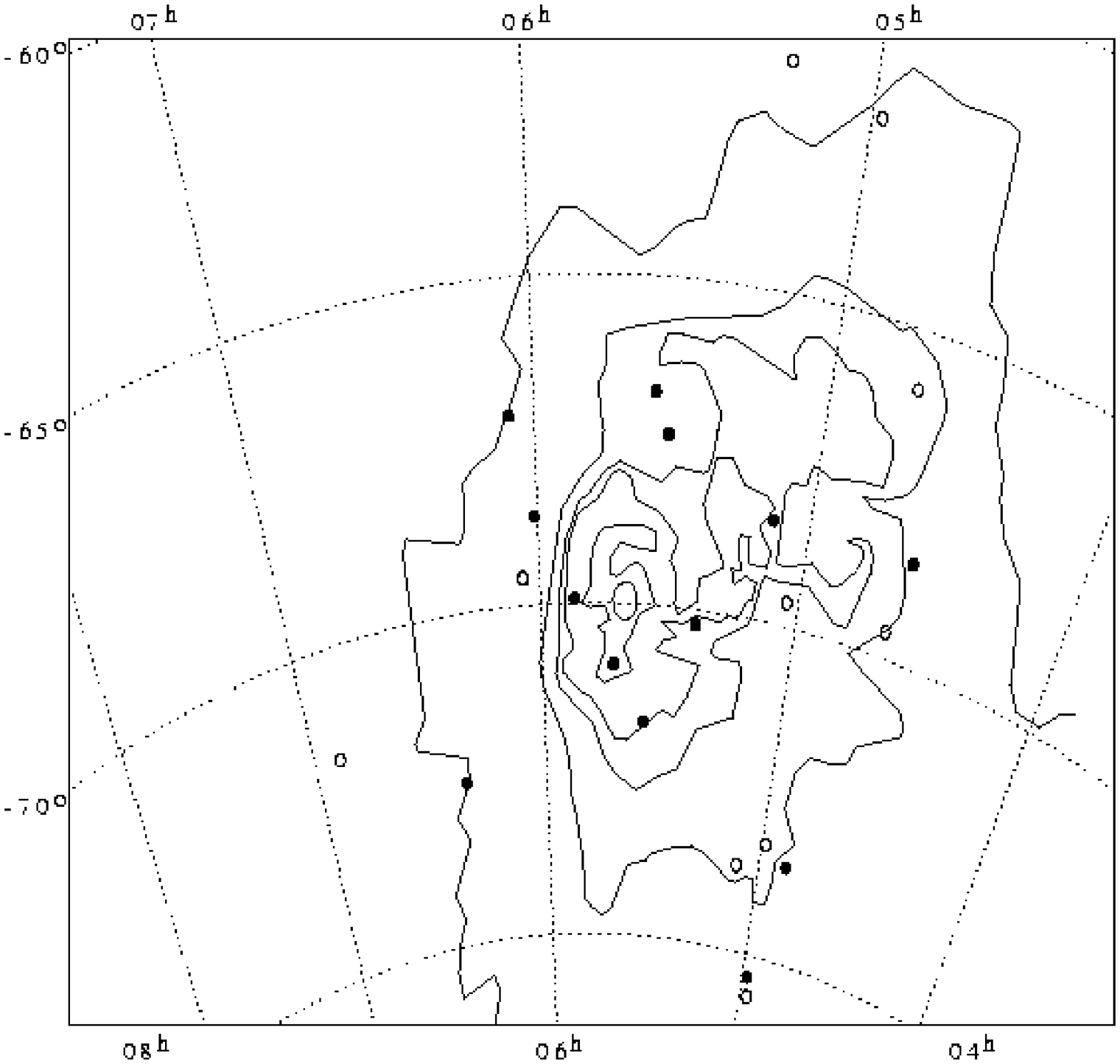}}
\caption{Angular distribution of the observed LMC background known galaxies (open circles) and candidate galaxies (filled circles). The solid lines represent H\,I contours of 5, 50, 100, 150, 400 and 600 in units {\rm $10^{19} atoms$  $cm^{-2}$} from Mathewson \& Ford (1984).}
\label{sample}
\end{figure*}

\begin{table*}
\tiny 
\caption{\scriptsize 
The observed objects towards the Magellanic Clouds. } 
\begin{tabular}{lccccccccl} 
\hline\hline 
Object&RA(2000)&Dec(2000)&$\ell$&b&B$_t$&Exp&V&V$_{lit}$&Type\\
&{\rm h:m:s~~~}&$^{\circ}$:$^{\prime}$~:$^{\prime\prime}~$&($^{\circ}$)&($^{\circ}$)&&(sec)&{\rm (km/s)}&{\rm (km/s)}&\\
\hline\hline 
&&&&&SMC main body&&&&\\
\hline
HS75-8&00:51:08&-73:39:22&302.96&-43.47&&4$\times$900&19650&--&E\\
AM0054-744sw&00:55:49&-74:30:58&302.53 &-42.61&&2$\times$900&10685&--&E\\
SMC-DEM92,AM0054-744ne&00:56:01&-74:30:40&302.52&-42.61&&3$\times$900&10369&--&E/S0\\
HS75-20&00:59:10&-74.02:39&302.20 &-43.07&&5$\times$900&19017&--&E\\
HS75-22&01:06:03&-73:59:52&301.55&-43.09&&2$\times$900&9890&--&E\\
HS75-23&01:06:07&-74:07:42&301.56 &-42.96&&5$\times$900&18950&--&E\\
HW60&01:09:27&-72:22:21&301.01&-44.69&&3$\times$900&17665&--&S\\
HS75-25,PMNJ0111-7302&01:11:33&-73:02:12&300.89&-44.01&&3$\times$900$+$600&19736&--&E\\
NGC643B,ESO29G53,IRAS01384-7515&01:39:14&-75:00:41&298.82 &-41.73&&2$\times$900&4006&3966&\\
\hline\hline 
&&&&&SMC surroundings&&&&\\
\hline
ESO28G12,IRAS00160-7325&00:18:20&-73:09:08&306.24&-43.76&14.90&2$\times$600&6200&6326&S0-a\\
HS75-10&00:52:34&-70:28:17&302.79&-46.66&&3$\times$900&18360&--&E\\
NGC406,ESO51G18,IRAS01057-7008&01:07:24&-69:52:35&300.91 &-47.19&13.02&3$\times$600&1391&1508&Sc\\
ESO52IG1-NED1&01:24:49&-68:37:21&298.37 &-48.21&14.76&2$\times$900&10778&11100&S0-a\\
NGC802,ESO52G13&01:59:06&-67:52:16&293.50 &-48.00&14.09&2$\times$900&1723&1504&SO-a\\
NGC813,ESO52G16&02:01:37&-68:26:21&293.52 &-47.38&13.78&600$+$420&8188&8160&S0-a\\
IC5339,ESO77G26,Fairall1051&23:38:05&-68:26:35&312.74 &-47.26&14.46&2$\times$900&12250&12328&E-SO\\
\hline\hline 
&&&&&SMC extended objects&&&&\\
\hline
HS75-13,H86-159&00:55:12&-72:40:57&302.54&-44.44&&3$\times$900&180&--&Star cluster\\
SMC-N63,HS75-18&00:58:17&-72:38:50&302.21&-44.47&&2$\times$900&132&--&H\,II Region\\
\hline\hline 
&&&&&LMC main body&&&&\\
\hline 
ESO55G33&04:38:51&-69:30:25&281.49 &-36.69&14.39&2$\times$900&5470&--&S0\\
NGC1669,ESO84G38&04:43:00&-65:48:53&276.96& -37.57&14.78&900$+$2$\times$600&5580&--&Sa\\
NGC1809,ESO56G48&05:02:05&-69:34:04&280.76 &-34.75&13.19&3$\times$900&1233&1301&Sc\\
ESO33G11&05:05:07&-73:39:08&285.39 &-33.34&14.35&2$\times$900&4550&--&SBaR\\
NEW GALAXY 1&05:07:38&-68:23:03&279.21 &-34.57&&2$\times$900$+$420&5782&--&\\
HS257,GSC916600034&05:22:45&-70:10:29&280.97 &-32.88&&2$\times$900&5560&--&\\
LMC-DEM225,IRAS05319-6723&05:31:49&-67:21:32&277.51 &-32.51&&3$\times$600&1376&--&\\
IRAS05338-6645&05:33:52&-66:43:18&276.74& -32.39&&3$\times$900&4320&--&\\
HS356,ESO56G154,KMHK1096,RXJ0534.0-7145&05:33:58&-71:45:20&282.62 &-31.70&&900$+$600&7150&7255&\\
HS394&05:42:02&-70:54:15&281.52 &-31.18&&3$\times$900&4567&--&\\
LMC-DEM329,IRAS05522-6952&05:51:42&-69:55:51&280.31 &-30.46&&2$\times$900&4380&--&\\
\hline\hline 
&&&&&LMC surroundings&&&&\\
\hline
HS17,RXSJ043612.5-682236,HP99-653&04:36:15&-68:22:10&280.27 &-37.33&&4$\times$900&19250&--&\\
HS45,IRASF04521-7333&04:51:09&-73:28:42&285.59 &-34.33&&3$\times$900&7565&--&\\
ESO33G2&04:55:59&-75:32:28&286.77& -33.29&14.66&2$\times$900&5408&5467&S0\\
ESO33G3&04:57:47&-73:13:51&285.11&-33.97&14.25&2$\times$600&7664&7677&E\\
NGC1765,ESO119G24&04:58:24&-62:01:41&271.83& -36.89&13.97&2$\times$900&8846&8758&E\\
ESO15G18&05:04:58&-81:18:38&293.89& -30.73&14.29&2$\times$900&4926&4903&E\\
ESO119G48&05:14:36&-61:28:54&270.78 &-35.09&13.51&2$\times$900&4549&4548&S0-a\\
HS449,GH90-060055-6840&06:00:43&-68:40:09&278.81 &-29.72&&4$\times$900&11305&--&\\
NGC2187A,ESO57G68sw,AM0604-693sw&06:03:44&-69:35:18&279.86&-29.44&12.94&2$\times$900&3769&3963&Sa\\
NGC2187B,ESO57G68ne,AM0604-693ne&06:03:52&-69:34:41&279.85& -29.43&13.16&2$\times$900&4519&4470&E\\
HS451&06:05:28&-67:07:10&277.03 &-29.28&&3$\times$900&7931&--&\\
SL887&06:21:01&-72:35:34&283.30 &-28.06&&$3\times$900$+$600&11466&--&\\
ESO58G19&06:52:57&-71:45:44&282.64 &-25.55&13.45&2$\times$900&4271&4251&S0-a\\
\hline\hline 
&&&&&LMC extended object&&&&\\
\hline
OHSC3,KMHK362&04:56:36&-75:14:29&287.42&-33.36&&4$\times$900&158&--&Star cluster\\
\hline\hline 
&&&&&Comparison B88&&&&\\
\hline
NGC1381&03:36:31&-35:17:39&236.467&-54.039&12.71&2$\times$600&1676&1776&S0\\
NGC1399&03:38:29&-35:26:58&236.714&-53.636&10.33&2$\times$900&1424&1434&E\\
NGC1411&03:38:45&-44:06:00&251.02&-52.52&12.18&2$\times$900&1100&1022&E-S0\\
NGC1404&03:38:52&-35:35:35&236.953&-53.555&10.89&2$\times$600&1970&1926&E\\
NGC1427&03:42:19&-35:23:37&236.598&-52.854&11.84&2$\times$600&1328&1425&E\\
NGC1600&04:31:40&-05:05:10&200.416&-33.242&12.04&2$\times$600&4703&4737&E\\
NGC6758&19:13:52&-56:18:33&340.573&-25.318&12.58&2$\times$600&3489&3408&E\\
IC4889&19:45:16&-54:20:37&343.538&-29.42&12.02&2$\times$600&2490&2521&E\\
IC1459&22:57:09&-36:27:37&4.665&-64.106&11.17&2$\times$600&1726&1679&E\\
\hline\hline
&&&&&South Polar Cap&&&&\\
\hline
NGC148&00:34:16&-31:47:10&340.648&-84.029&13.24&3$\times$600&1705&1516&S0\\
NGC155&00:34:40&-10:45:59&108.57&-73.169&14.28&5$\times$900&6101&6173&S0\\
NGC163&00:36:00&-10:07:17&110.121&-72.608&13.92&2$\times$900&5892&5981&E\\
NGC179&00:37:46&-17:50:56&103.462&-80.2&14.29&2$\times$900&6192&6006&E-S0\\
NGC277&00:51:17&-08:35:48&122.814&-71.468&14.73&900$+$600&4120&4327&E-SO\\ 
IC1633&01:09:55&-45:55:52&293.099&-70.843&12.55&2$\times$600&7437&7242&E\\
ESO476G4&01:21:07&-26:43:36&211.145&-83.373&13.86&2$\times$900&5922&5839&E-SOB\\
ESO352G55&01:21:33&-33:09:23&257.572&-81.133&14.63&2$\times$900&3747&3539&E-SO\\
ESO542G15&01:27:14&-21:46:24&181.527&-80.258&14.70&2$\times$900&5532&5567&SORing\\
NGC641&01:38:39&-42:31:40&273.994&-71.847&13.42&2$\times$900&6306&6454&E-SO\\
NGC720&01:53:00&-13:44:20&173.019&-70.358&11.40&2$\times$600&1563&1736&E\\
NGC7736&23:42:26&-19:27:09&55.15&-72.418&13.80&4$\times$900&4492&4511&S0\\
NGC7761&23:51:29&-13:22:53&74.423&-70.37&14.14&3$\times$900&7087&7082&SO\\
\hline
\end{tabular} 
\end{table*}
 
\subsection{South Polar Cap} 
 
The South Polar Cap sample (b $<$ -70$^{\circ}$) consists of 13 early type galaxies to minimize stellar population variations.
SFD98 pointed out the existence of low reddening regions between cirrus filaments near the Galactic Poles, and in some regions outside the Polar Caps as well at intermediate latitudes. Some of the latter regions have E(B-V)$_{FIR}$ values four times less than those estimated for the Polar Caps averaged over regions of diameter  10$^{\circ}$, which are E(B-V)$_{FIR}$ = 0.015 and 0.018, respectively for the Northern and Southern Polar Caps. We observed these galaxies to create reddening-free galaxy templates under the same observational conditions as the Magellanic Clouds sample. We also observed 9 intermediate galactic latitude galaxies (-64$^{\circ} < b < -29^{\circ}$) which are in common with those in B88's red stellar population templates in early type galaxies. The latter galaxies were observed for comparison purposes. The samples  are given in Table 1.
 
\section{Observations and reductions} 
 
The spectra were collected with the 2.15-meter telescope at the 
Complejo Astron\'omico El Leoncito (CASLEO, San Juan, Argentina) 
in December 1995 and October 1998. We employed a CCD camera 
attached to the REOSC spectrograph. The detector was a Tektronics 
chip of 1024$\times$1024 pixels of size 24{\rm $\mu \times 24 \mu$}. We 
used a grating of 300 grooves {\rm mm$^{-1}$} producing an average 
dispersion of $\approx$143{\rm \AA/mm} or 3.43{\rm \AA/pixel}. 
The spectral coverage was 3600 {\rm \AA} \ - 6800 {\rm \AA}. At least two 
exposures of each object were taken in order to correct for cosmic 
rays. The exposure times are given in Table 1. The  standard stars 
EG21 and LTT3864 (Baldwin \& Stone 1984) were observed for flux 
calibrations. He-Ar lamp exposures were taken following that of 
the object or standard star for wavelength calibrations. The slit 
width was 4$^{\prime\prime}$ providing a resolution (FWHM) of $\approx$ 14 {\rm \AA} \ 
from comparison lines. The slit was set in the E-W direction, and 
its length projected on the chip (4.7$^{\prime}$)  provided a wide range of 
pixel rows for background subtractions. 
 
The reductions were carried out at the Instituto de F\'{\i}sica, 
UFRGS (Porto Alegre - Brazil) and Observatorio Astron\'omico, 
Universidad Nacional de C\'ordoba (C\'ordoba - Argentina), with 
the {\bf IRAF} package following standard procedures. The galaxy 
spectra were extracted along the slit according to the dimensions of each galaxy nuclear region. Typical extractions were $\approx$ 8-10$^{\prime\prime}$. 
 
Since the spectral resolution was chosen for stellar population purposes, it is not ideal for velocity measurements. At any rate, we measured velocities and the agreement with values in common with the literature is good (Table 1). In one case, ESO28G12, the current LEDA and NED velocities are significantly different, and the present observation confirms the latter value (Table 1). The measured velocities  were used to bring the galaxy spectra to the rest frame, which is necessary for the subsequent stellar population analysis.

 Table 1 shows that the galaxies behind the SMC main body are mostly in the range 10000 $< V({\rm km/s}) <$ 20000, while those behind the LMC main body are mostly in the range 4000 $< V({\rm km/s}) <$ 6000. Since we have not a priori selected galaxies differently between the two Clouds the result suggests a real effect, in the sense that towards the LMC a closer galaxy ensemble occurs, which has no counterpart in the SMC background.

Figures 3 and 4 show the resulting rest-frame flux calibrated spectra for the Clouds background galaxies, for which the present study established or confirmed their nature (Sect. 4). Most of these galaxies have red stellar population nuclei, but some are bluer and present features denoting recent star formation such as Balmer absorption lines (e.g. HS75-20 in the SMC, Fig. 3) or emission lines typical of nuclear H\,II Regions (e.g. IRAS05538-6645 in the LMC, Fig. 4).

\section{Nature of the objects towards the Clouds}

We established the nature of the objects towards the Clouds  from the  radial velocity measurements (Table 1). In the present study velocities resulted either V $>$ 1000 {\rm km/s} or V $<$ 200 {\rm km/s}, undoubtedly characterizing background galaxies and Magellanic Clouds' internal objects, respectively. Most of the objects observed from  Hodge \& Snow's (1975) list are confirmed as galaxies, except HS75-13 and HS75-18 (upper panel of Fig. 5) which turned out to be a SMC star cluster (H86-159) and a SMC H\,II region (SMC-N63), respectively (see Sect. 4.1). OHSC3 is confirmed as a LMC star cluster (upper panel of Fig. 5). The objects SMC-DEM92 and HW60 in the SMC, and  LMC-DEM225, LMC-DEM329, HS17, HS45, HS257, HS356,  HS394, HS449, HS451 and SL887 in the LMC (originally classified as star clusters or H\,II regions) are shown to be galaxies. In the following we comment on the properties of some objects:

One observed galaxy  was not catalogued previously (New Galaxy 1, in Table 1).

HW60 appears to have a companion galaxy (New Galaxy 2, located at J2000  01$^h$09$^m$23$^s$ -72$^{\circ}$22$^{\prime}$15$^{\prime\prime}$) at $\approx$ 0.4$^{\prime}$ to the northwest with dimensions 0.2$^{\prime}$$\times$0.15$^{\prime}$. They possibly form an interacting system, which is supported by the fact that the HW60's spectrum is blue and presents strong emission lines, suggesting recent star formation (Fig. 3). Owing to the high radial velocity we do not have the H$_{\alpha}$, [NII] and [SII] region. Therefore, some nuclear activity cannot be ruled out.

SMC-DEM92 is the brighter member of the interacting pair AM0054-744 (Arp \& Madore 1987). We also observed the companion AM0054-744sw (Table 1, Fig. 3). Their radial velocities (Table 1)  are comparable, supporting an interaction. A similar case is the interacting pair NGC2187A and NGC2187B (Table 1), which is also an entry in  Arp \& Madore's catalogue.

In the SMC background the galaxy HS75-25 is an X-ray emitter  (Haberl et al. 2000), see present Table 1 for coordinates and other designation. Behind the LMC are X-ray emitters corresponding to HS17, HS257 (Haberl \& Pietsch 1999) and HS356 (Crampton et al. 1997). In particular, HS356 in a rich LMC field has been catalogued not only as star cluster (Hodge \& Sexton 1966, Kontizas et al. 1990) but as a galaxy as well (Lauberts 1982).

ESO52IG1 (Lauberts 1982) refers to a compact group of galaxies, studied by S\'ersic (1974) -- Se10/2,  and Arp \& Madore (1987) -- AM0123-685. The present galaxy is the brightest member of the group (accurate coordinates in Table 1). We point out that the literature coordinates often refer to the group centre. The NED database currently lists the 3 brighter members, but all coordinates are systematically shifted $\approx$ 0.7$^{\prime}$ to the northeast.

Some galaxies behind the Clouds are IRAS sources (Table 1).  

\begin{figure}
\resizebox{\hsize}{!}{\includegraphics{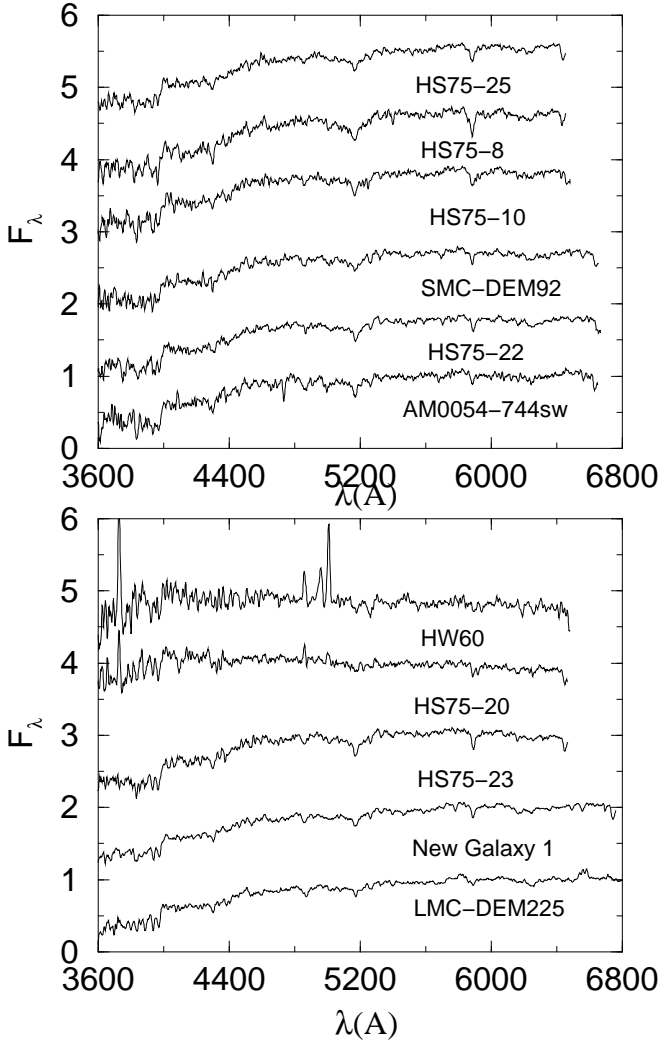}}
\caption{Rest-frame spectra of objects which turned out to be or were confirmed as SMC background galaxies, together with two LMC's (New Galaxy 1 and LMC-DEM225)}
\label{sample}
\end{figure}

\begin{figure}
\resizebox{\hsize}{!}{\includegraphics{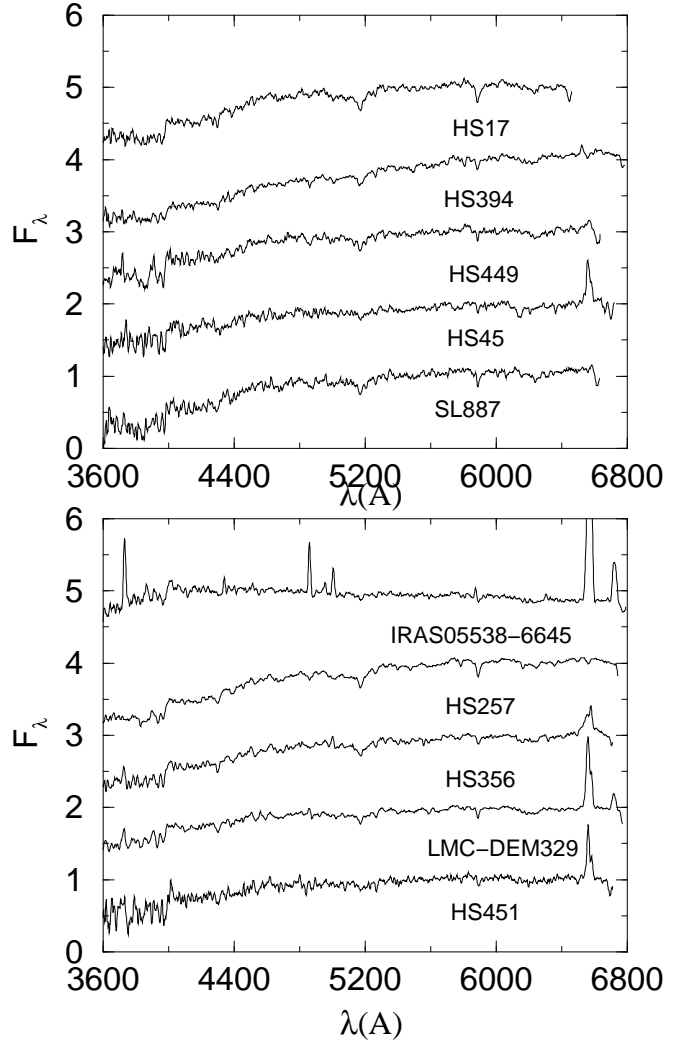}}
\caption{Same as Fig.3 for LMC background galaxies.}
\label{sample}
\end{figure}

\begin{figure}
\resizebox{\hsize}{!}{\includegraphics{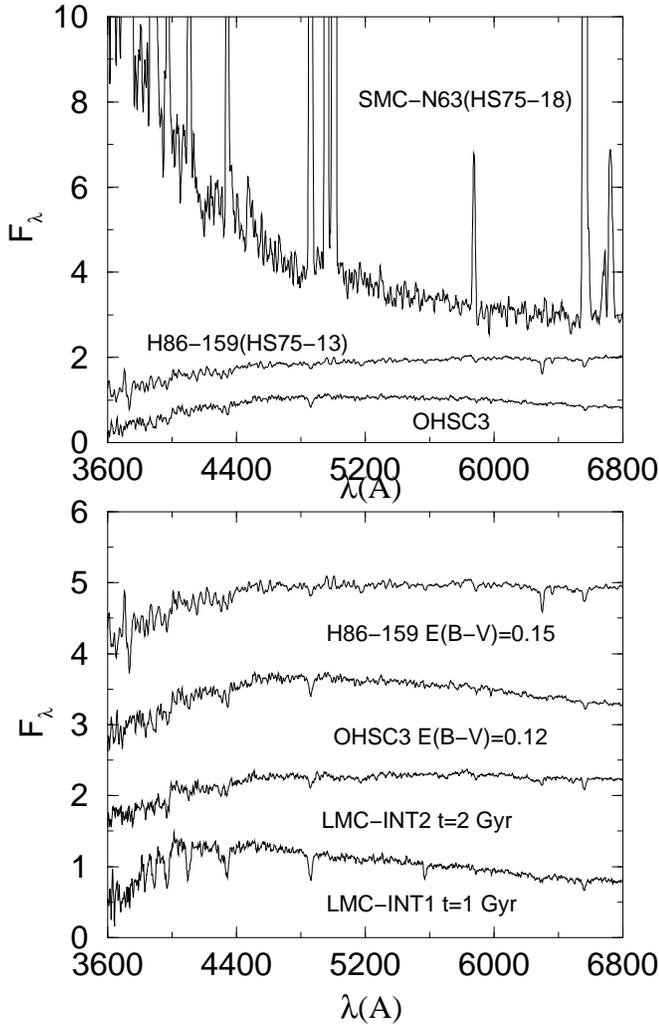}}
\caption{Upper panel: Objects which turned out to be or were confirmed as extended objects belonging to the SMC (H\,II Region HS75-18 and star cluster HS75-13), and LMC (star cluster OHSC3). Lower panel: age and reddening estimates using LMC cluster templates}
\label{sample}
\end{figure}

\subsection{Discussion of the LMC and SMC extended objects}

Bica \& Schmitt (1995) and Bica \& Dutra (2000) indicated that the object  SMC-N63 (Henize 1956) is also present in several other emission object catalogues as L61-331 (Lindsay 1961), SMC-DEM94 (Davies et al. 1976) and MA1065 (Meyssonnier \& Azzopardi 1993). The present cross-identification of this object with HS75-18 together with its CCD spectrum (upper panel of Fig. 5) conclusively establishes its nature as an H\,II region. The stellar content of the H\,II region was catalogued as the star cluster SMC-OGLE113 (Pietrzy\'nski et al. 1998). The gas reddening including the foreground reddening can be derived from the emission line spectrum (upper panel of Fig. 5) using the Balmer decrement F$_{H_{\alpha}}$/F$_{H_{\beta}}$ = 3.31. Assuming case B of the recombination-line theory the intrinsic ratio is  (F$_{H_{\alpha}}$/F$_{H_{\beta}}$)$_0$ = 2.87 (Osterbrock 1989), wherefrom  we derive E(B-V) = 0.13. This value is lower than SFD98's dust emission reddening estimate E(B-V)$_{FIR}$ = 0.37. A possible explanation for this difference is that the molecular cloud related to this recent star forming region is located behind SMC-N63.

Integrated spectra of star clusters as compared to template cluster spectra of different ages and metallicities can provide parameters such as age and reddening (e.g. Piatti et al. 1998, Ahumada et al. 2000). In the present study the signal-to-noise ratios of the H86-159 and OHSC3 spectra are not ideal for a detailed study of the cluster spectral features, but the spectral distribution can be compared to templates (lower panel of Fig. 5).

Using a deep plate from the CTIO 4-meter telescope Hodge (1986) catalogued the star cluster H86-159, and CCD images confirmed that (SMC-OGLE102 -- Pietrzy\'nski et al. 1998). The present study cross-identifies the latter object with HS75-13 and its spectrum (upper panel Fig. 5) confirms that one is not dealing with a galaxy. Recently de Oliveira et al. (2000) estimated a total reddening E(B-V)$_{MW+SMC}$ = 0.10 and age $\approx$ 500 {\rm Myr} for H86-159 from a colour-magnitude diagram (CMD) extracted from the OGLE photometric database (Udalski et al. 1998).  As comparison (lower panel of Fig. 5)  we show the  reddening-free LMC cluster templates LMC-Int1 and LMC-Int2, respectively of  ages 1 and 2 Gyr (Dutra et al. 1999). By applying reddening corrections to the spectrum of H86-159 following Seaton's (1979) extinction law, we derive  E(B-V)$_{MW+SMC}$ = 0.15 and an age of 2 {\rm Gyr}. In the cluster direction SFD98's dust emission reddening  is E(B-V)$_{FIR}$ = 0.38. The object is central in the SMC and star forming regions are nearby (traces of diffuse gas  are denoted by [OII] $\lambda$3727 {\rm \AA}, [OIII] $\lambda\lambda$ 4959,5007 {\rm \AA} \, superimposed on the cluster spectrum), so that important dust emission is expected in the area.  The high dust emission reddening as compared to the stellar content methods (CMD and spectrum) suggests the presence of dust in the cluster background within the SMC, similarly to star cluster directions in the Galaxy (Dutra \& Bica 2000). In contrast to background galaxies, Cloud clusters do not necessarily probe the total internal dust column of the Magellanic Clouds themselves. The age difference for the cluster can be explained by the fact that it is a poorly populated cluster. The turnoff is close to the photometric limit and the field is rich, which can affect significantly the CMD age determination. On the other hand, stochastic effects for bright stars and field contamination for such low mass clusters may cause important uncertainties on the integrated properties (e.g. Geisler et al. 1997). At any rate the bracket 0.5 - 2 {\rm Gyr} is a considerable constraint on the cluster age.

OHSC3 (Olszewski et al. 1988) or KMHK362 (Kontizas et al. 1990) was catalogued as a star cluster. It called our attention owing to the compact nature in ESO/SERC sky survey plates. There is a neighbouring galaxy $\approx$ 2.3$^{\prime}$ to the northeast. This galaxy (New Galaxy 3) is an inclined spiral with dimensions $\approx$ 0.6$^{\prime}$$\times$0.2$^{\prime}$ located at J2000  04$^h$57$^m$05$^s$ -75$^{\circ}$13$^{\prime}$08$^{\prime\prime}$. We suspected a galaxy pair, but this is not confirmed by the OHSC3 spectrum. In this direction SFD98 predict E(B-V)$_{FIR}$ = 0.12, which was applied in the reddening correction of the OHSC3 spectrum (lower panel of Fig. 5). The spectral properties appear to be intermediate between the two template spectra leading to an age $\approx$ 1.5 {\rm Gyr}.

\section{Reddening-free galaxy templates}

Red stellar population galaxy nuclei are ideal as reddening probes since the spectral distribution is essentially insensitive to age variations of the components, and present a small dependence on metallicity (B88). Blue stellar populations have stronger continuum variations with age distribution of the components. 
 
B88 studied the stellar populations of early and late type galaxies by means of their nuclear integrated spectra. Early and late type galaxy nuclei were studied separately considering also luminosity differences (B88 and references therein).  Spectra  with similar equivalent widths and dereddened continuum distribution were grouped into high signal-to-noise templates. These templates represent the most frequent types of stellar populations found in normal galaxy nuclei. The  early type templates E1 to E3 represent a decreasing metallicity sequence among giant early type galaxy nuclei dominated by old (red) stellar populations. E7 represents nuclei dominated by old populations but with significant contribution of 0.5 to 1 Gyr components. S1 to S3 represent a  similar metallicity sequence to E1-E3  for red stellar populations occurring in giant spiral galaxy nuclei. Finally, S4 to S7 is a sequence of giant spiral galaxy nuclei with increasing contributions of young stellar populations.

For the spectral comparisons between sample galaxies and templates we employ equivalent widths (W) of strong absorption features. We use as metal features  K CaII, CN, G Band, MgI and NaI, together with four Balmer lines. In Table 2 are shown W values for templates and individual galaxies measured with continuum tracings and feature windows following Bica \& Alloin (1986) and Bica et al. (1994).  For bluer stellar populations we increased the resolution of spectral properties by creating the intermediate templates S5/S6 (0.5$\times$S5 + 0.5$\times$S6), S6/S7A (0.5$\times$S6 + 0.5$\times$S7) and S6/S7B (0.25$\times$S6 + 0.75$\times$S7). This procedure in turn provides a higher resolution in reddening determinations since the continuum varies strongly for increasing contents of younger populations in galaxy nuclei. Typical W errors are $\approx$ 5 \% and depend mostly on signal-to-noise ratio and uncertainties in the continuum positioning.

Red stellar populations are the most frequent types in the background samples (Figs. 3 and 4). In addition to the high reddening accuracy that they can provide (Sect. 6), it is important also to minimize possible observational uncertainties by creating templates from galaxies observed in the same observing runs. For these purposes we built new red stellar population templates using South Polar Cap galaxies (Sect. 2.2) and some galaxies in common with B88, which have E(B-V)$_{FIR}<$ 0.02 (Table 3) to avoid dust cirrus (Sect. 2.2). Measurements of Ws for these galaxies (Table 2) allowed us to classify them into types E1, E2 or E3.  We dereddened the spectra using E(B-V)$_{FIR}$ values and Seaton's (1979) galactic extinction law. At this stage, the E2 and E3 member galaxies turned out to be  very similar in terms of continuum distribution.  Therefore we adopted two reference spectra T1 and T23 as counterparts in the present study of the E1 and E2/E3 templates. Members of T1 and T23 are indicated in Table 3. The remaining galaxies with somewhat higher reddening values (Table 3) have either internal reddening as dusty ellipticals (Ferrari et al. 1999) and/or foreground contribution. It is possible to determine a spectroscopic reddening value E(B-V) by fitting the observed galaxy spectrum to that of the corresponding template with similar Ws and applying continuum corrections according to Seaton's law. We provide in column 4 of Table 3 results for individual galaxies in the templates T1 and T23 themselves and additional red stellar population galaxies in the same regions.

\begin{table*} 
\tiny
\caption{\scriptsize 
Ws for strong absorption features in the template and individual galaxy spectra}
\centerline{ 
\begin{tabular}{lccccccccc} 
\hline\hline 
Object & K & H$_\delta $ & CN & G & H$\gamma $ & H$_\beta $ & MgI & NaI 
& H$_\alpha $ \\
Windows & 3908-3952 & 4082-4124 & 4150-4214 & 4284-4318 & 4318-4364 & 
4846-4884 & 5156-5196 & 5880-5914 & 6540-6586   \\
\hline
&&&&&B88 Templates&&&&\\
\hline
E1&17.5&5.7&15.1&9.6&5.2&3.7&10.9&6.6&0.2\\
E2&17.9&5.8&12.1&9.6&5.7&3.5&9.6&5.4&1.6 \\
E3&16.1&2.3&7.9&8.9&4.3&4.2&8.0&4.2&2.2\\
E7&12.3&6.1&6.9&6.9&5.4&4.9&7.4&4.1&------\\
S4&13.2&4.7&5.8&7.3&5.0&3.5&6.1&4.1&e\\
S5&7.7&3.9&5.5&6.3&3.3&0.9&6.1&4.1&e\\
S5/S6&6.1&4.1&4.0&5.5&3.5&0.0&5.3&3.7&e\\
S6&5.6&4.2&2.6&4.7&3.6&e&4.0&3.4&e\\
S6/S7A&2.5&4.7&2.0&3.5&3.6&e&3.7&3.2&e\\
S6/S7B&3.1&4.9&2.0&3.3&4.0&e&3.8&3.3&e\\
S7&2.9&5.4&2.3&3.1&4.1&e&3.9&3.5&e\\
\hline
&&&&&New Red Population Templates&&&&\\
\hline
T1&18.6&6.1&15.0&9.6&4.3&3.4&11.3&7.0&0.2\\
T23&17.1&4.3&10.2&9.2&4.4&3.5&8.7&4.3&0.8 \\
\hline\hline
&&&&&Comparison (B88)&&&&\\
\hline 
NGC1381&18.4&3.6&10.0&8.7&4.4&4.0&8.2&4.2&1.9\\
NGC1399&19.4&7.5&16.7&8.6&4.3&4.1&12.6&7.5&1.6\\ 
NGC1411&16.7&4.3&10.6&9.4&4.3&3.2&8.4&4.7&--\\
NGC1404&18.7&7.2&12.7&11.0&4.4&3.4&10.1&6.7&2.3\\
NGC1427&19.0&5.5&11.8&9.8&6.4&4.1&9.5&4.5&1.1\\
NGC1600&18.5&6.4&15.5&8.9&4.3&4.1&12.1&5.8&2.2\\
NGC6758&18.7&8.6&15.3&10.7&5.6&4.2&10.1&6.1&1.4\\ 
IC4889 & 20.1 & 9.2 & 12.6 & 11.9 & 8.2 & 2.9 & 7.1 & 2.9 & --\\ 
IC1459&20.4&5.4&13.8&9.0&3.6&2.0&11.2&6.9&--\\ 
\hline
 &&&&&South Polar Cap&&&&\\ 
\hline 
NGC148&19.4&7.6&12.2&10.5&5.4&4.3&9.2&6.1&2.9\\
NGC155&17.3&4.8&10.7&10.7&6.1&5.4&9.5&4.2&--\\
NGC163 & 17.2 & 4.9 & 13.5 & 9.1 & 5.2 & 2.8 & 9.8 & 5.0 & 1.3 \\
NGC179&16.8&6.9&12.1&9.7&5.5&4.4&8.4&4.0&--\\
NGC277&17.9&6.4&13.3&10.7&6.1&4.0&10.1&5.2&--\\
IC1633 & 18.6 & 4.9 & 9.3 & 9.8 & 7.1 & 2.9 & 8.0 & 4.7 & -- \\ 
ESO476G4&16.2&2.4&8.0&9.3&3.5&2.7&8.0&3.8&2.0\\
ESO352G55&19.6&5.8&11.8&8.9&4.9&6.4&9.5&5.4&2.3\\ 
ESO542G15&17.0&4.3&9.5&8.8&5.1&3.9&8.2&4.1&1.6\\
NGC641&17.8&5.2&11.4&9.3&3.9&3.2&9.4&4.0&--\\
NGC720&18.7&7.1&15.3&9.4&4.9&2.2&10.9&6.9&--\\
NGC7736 & 19.0& 6.5&13.8&9.7&6.1&4.8&10.3&5.6&2.2\\ 
NGC7761&13.6&5.7&10.0&8.9&6.4&4.5&8.4&3.2&--\\
\hline
&&&&&SMC Main body&&&&\\
\hline
HS75-8 & 17.2 & 4.9 & 13.5 & 9.1 & 5.2 & 2.8 & 9.8 & 5.0 & 1.3 \\
AM0054-744sw & 20.4 & 1.7 & 6.7 & 9.2 & 3.9 & 5.1 & 8.5 & 4.1 & 2.3 \\
SMC-DEM92 & 16.7 & 1.3 & 9.2 & 8.1 & 5.9 & 3.6 & 6.9 & 3.6 & 1.8 \\
HS75-20 & 4.7 & 6.3 & 2.3 & 4.4 & 5.6 & e & 5.5 & 4.3 & ------  \\
HS75-22 & 16.5 & 4.5 & 12.9 & 8.8 & 5.1 & 2.4 & 8.9 & 4.5 & 1.0  \\
HS75-23 & 15.2 & ------ & 6.9 & 8.5 & 4.6 & 3.3 & 10.0 & 5.7 & ------ \\
HS75-25 & 18.6 & 4.9 & 9.3 & 9.8 & 7.1 & 2.9 & 8.0 & 4.7 & ------ \\ 
NGC643B & 7.9 & 8.4 & 4.6 & 4.9 & 7.6 & 2.9 & 3.7 & 2.4 & e \\
\hline
&&&&&SMC surroundings&&&&\\
\hline 
ESO28G12 & 6.5 & 4.8 & 2.9 & 6.3 & 3.9 & e & 4.7 & 3.2 & e \\ 
HS75-10 & 19.8 & 3.9 & 14.0 & 9.0 & 6.4 & 2.9 & 8.7 & 5.9 & 2.4\\ 
NGC406 & 9.8 & 5.0 & 2.3 & 4.1 & 5.7 & e & 3.3 & 1.7 & e  \\
ESO52IG1-NED1 & 18.5 & 4.9 & 13.3 & 9.1 & 4.8 & 4.0 & 9.5 & 5.1 & 2.6 \\
NGC802 & 0.7 & 6.0 & 1.5 & 2.5 & 5.0 & e & 2.3 & 2.1 & e \\ 
NGC813 & 14.3 & 3.7 & 8.7 & 8.8 & 5.2 & 3.5 & 6.8 & 3.1 & 2.8 \\ 
IC5339& 16.7 & 2.9 & 7.9 & 8.4 & 3.9 & 3.8 & 9.9 & 6.2 & ------ \\ 
\hline
&&&&&LMC&&&&\\
\hline\hline
&&&&&LMC Main body&&&&\\
\hline
ESO55G33 & 11.4 &  5.0 & 8.1 & 7.3 & 5.7 & 1.3 & 6.1 & 3.2 & e\\
NGC1669 & 5.6 & 5.2 & 2.5 & 4.7 & 3.5 & 2.3 & 6.0 & 2.8 & e\\
NGC1809 & 5.2 & 9.3 & 3.7 & 3.1 & 5.8 & 1.7 & 3.0 & 2.8 &e \\
ESO33G11 & 17.4 & 4.2 & 11.0 & 9.0 & 5.0 & 2.7 & 7.6 & 3.1 & e \\
NEW GALAXY 1 & 14.9 & 5.8 & 10.3 & 8.5 & 5.7 & 3.5 & 7.8 & 4.6 & 2.3 \\
HS257 & 20.2 & 6.4 & 13.3 & 9.8 & 4.6 & 3.5 & 10.0 & 5.5 & 1.9 \\ 
LMC-DEM225&12.1&4.2&6.2&6.8&5.0&4.3&5.7&3.6&------\\
IRAS05338-6645 & 4.9 & 4.6 & 3.5 & 3.6 & 1.4 & e & 3.0 & 2.0 &e\\
HS356 & 14.2 & 5.0 & 8.9 & 7.9 & 4.3 & 1.9 & 7.2 & 3.0 & e\\
HS394 & 16.6 & 3.4 & 8.1 & 9.7 & 5.5 & 4.1 & 8.2 & 3.4 & 1.3 \\
LMC-DEM329 & 8.0 & 4.4 & 6.4 & 5.8 & 3.0 & e & 5.5 & 2.8 & e \\
\hline\hline
&&&&&LMC surroundings&&&&\\
\hline
HS17 & 16.2 & 5.1 & 12.4 & 8.5 & 3.6 & 2.1 & 10.0 & 6.6 & ------ \\
HS45 & 10.7 & 6.8 & 5.1 & 7.3 & 7.8 & 2.8 & 5.3 & 2.5 & e \\
ESO33G2 & 14.0 & 3.4 & 9.5 & 8.0 & 3.4 & e & 6.3 & 3.5 & e \\
ESO33G3 &18.0&7.6&13.14&9.9&4.1&3.6&8.5&3.9&------\\
NGC1765 & 19.2 & 6.2 & 14.7 & 9.7 & 5.0 & 3.1 & 9.8 & 6.0 & 1.5 \\
ESO15G18 & 16.7 & 2.1 & 10.2 & 8.9 & 4.8 & 2.9 & 7.8 & 4.6 &e \\
ESO119G48 & 15.4 & 3.3 & 9.6 & 8.3 & 4.3 & 3.6 & 7.8 & 4.3 & 1.4 \\
HS449 & 13.5 & 4.0 & 8.4 & 7.7 & 6.4 & 4.0 & 7.8 & 3.8 &e \\
NGC2187A & 18.6 & 4.5 & 13.6 & 10.0 & 5.1 & 3.7 & 10.0 & 5.6 & 2.3 \\
HS451 & 8.9 & 3.5 & 6.9 & 6.4 & 4.3 & 2.6 & 4.3 & 2.8 & e \\
SL887 & 17.2 & 6.0 & 13.3 & 9.1 & 3.7 & 4.9 & 8.1 & 4.7 & ------ \\ 
ESO58G19 & 15.7 & 3.0 & 11.9 & 10.3 & 5.1 & 3.0 & 8.4 & 4.5 & 1.4 \\ 
\hline
\end{tabular}}
\begin{list}{}
\item  Note: e - indicates Balmer line in emission. 
\end{list}
\end{table*} 

Table 3  indicates that dust emission and spectroscopic reddening values are consistent for small amounts of reddening. The average of spectroscopic reddening values in the present South Polar Cap sample (13 galaxies) is  E(B-V) = 0.02$\pm$ 0.01, which is consistent  with SFD98's estimate (Sect. 2.2).

\begin{table*} 
\caption{\scriptsize 
Spectroscopic and dust emission reddening values for observed galaxies in the B88 and South Polar Cap samples.} 
\begin{tabular}{lcccccc} 
\hline\hline 
&&Comparison galaxies&from B88&\\
\hline
Object&Template&comments&E(B-V)&E(B-V)$_{FIR}$\\
\hline
NGC1381&T23&member&0.01&0.01\\
NGC1399&T1&member&0.01&0.01\\
NGC1411&T23&member&0.01&0.01\\
NGC1404&T1&member&0.02&0.01\\
NGC1427&T23&member&0.01&0.01\\
NGC1600&T1&dusty&0.02&0.04\\
NGC6758&T1&dusty&0.05&0.07\\ 
IC4889 &T23&dusty&0.05&0.05\\ 
IC1459&T1&member&0.03&0.02\\ 
\hline 
&&South Polar Cap&&\\ 
\hline 
Object&Template&Comments&E(B-V)&E(B-V)$_{FIR}$\\ 
\hline
NGC148&T23&member&0.02&0.02\\
NGC155&T23&&0.02&0.03\\
NGC163 &T1&&0.01&0.03\\
NGC179&T23&member&0.02&0.02\\
NGC277&T1&&0.01&0.04\\
IC1633 &T23&member&0.02&0.01\\
ESO476G4&T23&member&0.01&0.01\\
ESO352G55&T23&&0.05&0.03\\
ESO542G15&T23&member&0.01&0.02\\
NGC641&T23&member&0.02&0.02\\
NGC720&T1&member&0.01&0.02\\
NGC7736 &T1&&0.03&0.03\\ 
NGC7761&T23&&0.03&0.03\\ 
\hline
\end{tabular}
\end{table*}  

\section{Reddening towards the Magellanic Clouds}

For each background galaxy we searched for the template spectrum with comparable Ws (Table 2) which was assumed as the reddening-free reference stellar population. The resulting templates are shown in column 2 of Table 4. Making use of Seaton's law and varying the reddening amount we dereddened the observed galaxy spectrum to match  the template continuum distribution. The upper panel of Figure 6  illustrates the  reddening determination for a red stellar population galaxy (HS394). Note the important reddening effect in the observed spectrum. The lower panel illustrates a blue stellar population nucleus (IRAS05338-6645). The reddening values obtained by this spectroscopic procedure are given in column 3 of Table 4. For comparison purposes the dust emission reddening E(B-V)$_{FIR}$ value is shown in column 4. Finally, in column 5 we give the  H\,I column density in units of 10{\rm $^{19}$ atoms cm$^{-2}$} (Mathewson \& Ford 1984).

The uncertainties in the matching of the continuum distributions is small $\epsilon$E(B-V) $\approx$ 0.01. The largest source of uncertainties arises from the determination of the stellar population. For red stellar populations this uncertainty is typically $\epsilon$E(B-V) $\approx$ 0.02, and somewhat larger for blue stellar populations ($\epsilon$E(B-V) $\approx$ 0.05). 

Out of 36 SMC and LMC background galaxies, 31 show good agreement ($\delta$E(B-V) $\leq$ 0.10) between the spectroscopic and dust emission reddening values (Table 4), leading to a r.m.s. of differences of 0.04. One case of significant difference (HS75-8) is behind the SMC main body, and the remaining ones (New Galaxy 1, HS356, HS394 and LMC-DEM329) behind the LMC main body. In the latter 5 cases E(B-V)$_{FIR}$ is larger and H\,I column densities are important (Table 4). Since the slit apertures are typically 4$^{\prime\prime}$ $\times$ 8-10$^{\prime\prime}$ (Sect. 3) and SFD98's pixel dimensions  142$^{\prime\prime}$ $\times$ 142$^{\prime\prime}$, a possible explanation is that the dust distribution is patchy with a scale significant smaller than SFD98's pixel. We note that out of the five galaxies with significant differences three have red stellar populations (Table 4) and the spectroscopic reddening values are accurate. 

The largest difference occurs for New Galaxy 1 which has a very high reddening value E(B-V)$_{FIR}$ = 0.68 (Table 4). This galaxy is projected at the edge of the shell emission nebula LMC-DEM76 (Davies et al. 1976). An alternative explanation to dust small scale variations is that dust in this region may be heated beyond the temperature correction range employed by SFD98. Evidence of a similar effect was pointed out by Dutra \& Bica (2000) for the Galactic Center direction.

The galaxy LMC-DEM225 (Fig. 3) has a rare stellar population type dominated by a red stellar population, but with an important Balmer-line in absorption component in the range 3700 {\rm \AA} \, to 3900 {\rm \AA}. The most similar template in terms of Ws available in B88 is E7 (Table 2). The enhanced Balmer lines in the violet region might be explained by a somewhat younger burst (100 - 500 {\rm Myr}) in LMC-DEM225, rather than the 1 {\rm Gyr} burst in E7 (B88). Owing to the stellar population difference we have not determined reddening for this galaxy.

\begin{figure}
\resizebox{\hsize}{!}{\includegraphics{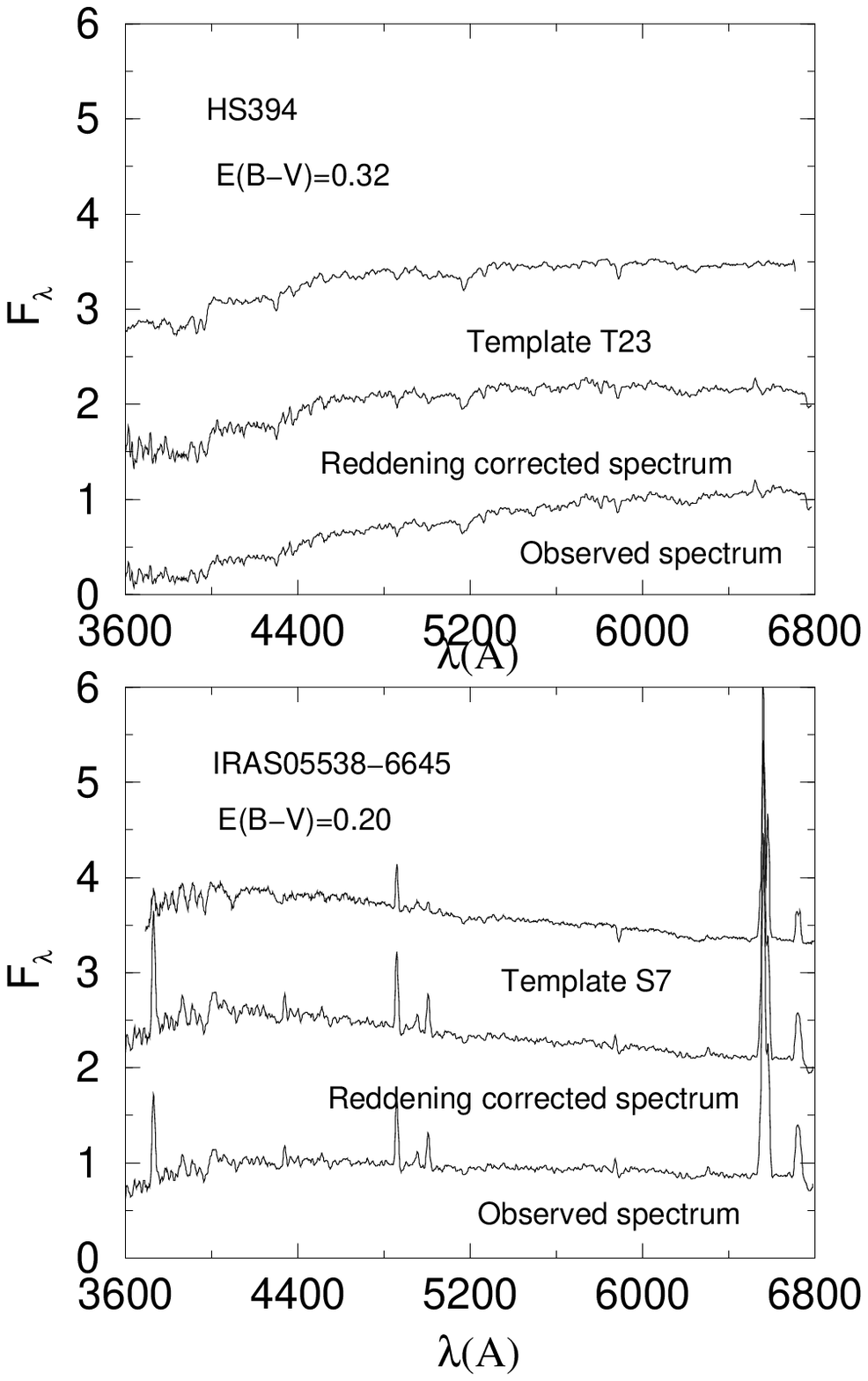}}
\caption{Upper panel: Example of reddening determination for a red stellar population nucleus. Lower panel: Example of reddening determination for a blue stellar population nucleus.}
\label{sample}
\end{figure}

The galaxies projected on SMC surroundings (H\,I column density $<$50$\times$10{\rm $^{19}$ atoms cm$^{-2}$})  have an average spectroscopic reddening value of E(B-V) = 0.01$\pm$ 0.02. This value is comparable to that of the South Polar Cap galaxies (Sect. 5), showing that at least in the direction of the SMC no significant difference occurs between the Polar region  and b $\approx$ -45$^{\circ}$. This value is  also comparable to the average dust emission reddening in the SMC surroundings E(B-V)$_{FIR}$ = 0.03$\pm$ 0.01 (Table 4) and that derived by SFD98 in the same region (Sect. 1). Considering the  galaxies behind the SMC main body the average spectroscopic reddening is E(B-V) = 0.05$\pm$ 0.05 and that from dust emission in the corresponding lines of sight is E(B-V)$_{FIR}$ = 0.07$\pm$ 0.04. By subtracting the average of reddening values in the main body directions from those in the surroundings, both the spectroscopic and dust emission  methods predict an internal SMC reddening $\Delta$E(B-V) = 0.04 which is consistent with the fact that important H\,I column density differences occur between these regions (Table 4). 

In the LMC surroundings (HI density column $<$50$\times${\rm 10$^{19}$ atoms cm$^{-2}$}) the average spectroscopic reddening is E(B-V) = 0.06$\pm$ 0.03 which is similar to the average dust emission reddening value in the corresponding directions E(B-V)$_{FIR}$ = 0.08$\pm$ 0.04 (Table 4). These values are in turn similar to those derived by SFD98 (Sect. 1). We conclude that in the LMC direction at b $\approx$ -33$^{\circ}$ the Milky Way reddening becomes evident as compared to the Polar Cap. For the LMC main body the average spectroscopic reddening is E(B-V) = 0.12$\pm$ 0.10. Considering only the red population galaxies, which provide the most accurate determinations, the result is basically the same E(B-V) = 0.15$\pm$ 0.11. This suggests an intrinsic dispersion among lines of sight. We estimate $\Delta$E(B-V) = 0.06 between LMC main body directions and surroundings.

\begin{table*}  
\caption{\scriptsize 
Spectral and dust emission reddening values for galaxies towards the Magellanic Clouds.} 
\begin{tabular}{lcccccc} 
\hline\hline
&&SMC&&\\
\hline 
Object&Template&E(B-V)&E(B-V)$_{FIR}$&HI column\\ 
\hline
&&Main body&&\\
\hline
HS75-8&T1&0.03&0.15&675\\
AM0054-744sw&T23&0.00&0.05&136\\
SMC-DEM92&T23&0.00&0.05&130\\
HS75-20&S6/S7B&0.10&0.05&210\\
HS75-22&T23&0.04&0.04&140\\
HS75-23&T23&0.00&0.04&123\\
HS75-25&T23&0.10&0.10&484\\
NGC643B&S6&0.12&0.05&82\\
\hline
&&surroundings&&\\
\hline
ESO28G12&S6&0.00&0.04&37\\
HS75-10&T1&0.01&0.02&27\\
NGC406&S6/S7A&0.00&0.02&18\\
ESO52IG1-NED1&T23&0.04&0.03&20\\
NGC802&S7&0.00&0.02&28\\
IC5339&T23&0.01&0.04&outside\\
\hline\hline
&&LMC&&\\
\hline 
Object&Template&E(B-V)&E(B-V)$_{FIR}$&HI column\\ 
\hline
&&Main body&&\\
\hline
ESO55G33&S5&0.10&0.15&50\\
NGC1669&S6&0.10&0.05&75\\
NGC1809&S7&0.23&0.24&75\\
ESO33G11&T23&0.09&0.13&55\\
NEW GALAXY 1&T23&0.04&0.68&136\\
HS257&T23&0.15&0.24&141\\
IRAS05538-6645&S7&0.20&0.13&87\\
HS356&S4&0.00&0.13&158\\
HS394&T23&0.32&0.50&446\\
LMC-DEM329&S56&0.02&0.22&186\\
\hline
&&surroundings&&\\
\hline
HS17&T23&0.06&0.06&40\\
HS45&S5&0.02&0.12&44\\
ESO33G2&S4&0.08&0.13&outside\\
ESO33G3&T1&0.09&0.11&outside\\
NGC1765&T1&0.03&0.03&7\\
ESO15G18&T23&0.10&0.10&outside\\
ESO119G48&T23&0.01&0.02&outside\\
HS449&T23&0.01&0.06&25\\
NGC2187A&T1&0.07&0.10&27\\
HS451&S6&0.05&0.07&5\\
SL887&T23&0.07&0.08&outside\\
ESO58G19&T23&0.10&0.11&outside\\
\hline 
\end{tabular}
\end{table*}

\section{Concluding remarks}

Spectra of nuclear regions of galaxies towards the Magellanic Clouds are useful reddening probes in the Milky Way and the Clouds themselves. By means of radial velocities we established the nature of $\approx$ 20 objects in terms of background galaxies or star clusters and H\,II regions belonging to the Clouds. The radial velocities of the observed galaxies behind the Magellanic Clouds reveal an interesting spatial distribution: most of the brighter galaxies in the main body of the LMC are in the range 4000 $< V({\rm km/s}) <$ 6000, while for the SMC they are more distant in the range 10000 $< V({\rm km/s}) <$ 20000. 

We estimated reddening values by comparing observed galaxy spectra with reddening-free templates of similar stellar populations. We inferred the reddening distribution throughtout the Clouds by means of 18 galaxies in the main bodies and 18 in the surroundings. The SMC foreground reddening (E(B-V)$_{MW}$ = 0.01) is comparable to that of the South Polar Cap, while that of the  LMC, at a lower galactic latitude is higher (E(B-V)$_{MW}$ = 0.06). Both Clouds are quite transparent, at least in the sampled lines of sight. The  derived average internal reddening values are E(B-V)$_i$ = 0.04 and 0.06, respectively for the SMC and LMC. The resulting statistics on reddening values are likely to be skewed toward lower reddening values, since heavily reddened galaxies are less likely to be detected, especially in the most crowded central regions in the Clouds.

Following the reddening study in the Milky Way by Dutra \& Bica (2000), which analyses reddening values derived from dust emission and stellar content of star clusters, the present method probes the foreground and internal reddening in the Clouds by means of background galaxies. We conclude that the dust emission and stellar population reddening values towards the Clouds agree well for 86 \% of the present sample. The significant differences found are in the sense of larger E(B-V)$_{FIR}$ values. Possible explanations are patchy dust distribution in a scale smaller than SFD98's resolution or dust heated above the range of the temperature map.

In order to map out in detail the dust distribution in the Magellanic Clouds large samples of background galaxies are necessary. We note that several of the present main body galaxies are not obvious on Sky Survey images, because of  crowding and proximity to extended objects in the Clouds. A fundamental question is whether more reddened zones exist in the Clouds, especially in the LMC. It would be important to carry out higher resolution imaging to identify galaxies for spectroscopy in large telescopes.

\section*{Acknowledgments} 
 
We thank the CASLEO staff for hospitality and support during the observing runs. The authors acknowledge use of the CCD and data acquisition system supported under U.S. National Science Foundation grant AST-90-15827 to R.M. Rich. 
We have made use of the LEDA database, ({\it http://leda.univ-lyon1.fr}), and the NASA/IPAC Extragalactic Database (NED) which is operated by the Jet Propulsion Laboratory, California Institute of Technology, under contract with the National Aeronautics and Space Administration. We use images from the Digitized Sky Survey (produced at the Space Telescope Science Institute under U.S. Government grant NAG W-2166) by means of the Canadian Astronomy Data Centre (CADC) interface. This work was partially supported by the Brazilian institutions CNPq and FINEP, the Argentine institutions CONICET, ANPCyT and SECYT (UNC), and the VITAE and Antorchas foundations.

\end{document}